\documentstyle[twoside,fleqn,espcrc2,psfig]{article}

\newcommand{\AmS}{{\protect\the\textfont2
  A\kern-.1667em\lower.5ex\hbox{M}\kern-.125emS}}

\title{Extreme synchrotron blazars: the case of Mkn 501 
}

\author{Gabriele Ghisellini \address{Brera Astronomical Observatory,
        V. Bianchi, 46 Merate, Italy}
}
       
\begin{document}

\begin{abstract}
BeppoSAX observations of Mkn501 in April 1997 (Pian et al. 1998), 
have revealed an
extraordinary X--ray emission from this BL Lac object, during a phase 
of high activity at TeV energies, as monitored with the Whipple, HEGRA 
and CAT Cerenkov telescopes.  
The 0.1--200 keV spectrum was hard, and the X--ray power output peaked at 
or above 100 keV, 2 or 3 orders of magnitude more than what indicated
by previuos observations, while the luminosity increased by at least a
factor $\sim$20.
The X--ray spectrum hardens when the source is brighter, but 
variations seem limited to energies greater than $\sim$0.5 keV.
All these unprecedented spectral information pose severe constraints
to all models, and we discuss in particular how the homogenous, one--zone 
synchrotron self Compton model must be modified to account for
the observed properties.
Other sources, besides Mkn 501, could undergo similar flares.

\end{abstract}

\maketitle

\section{INTRODUCTION}
BL Lac objects come in at least two flavours, according to their overall
spectral energy distribution (SED) viewed in a $\nu$--$\nu F_{\nu}$
representation: sources in the first group have the peak of their 
synchrotron emission in the IR--optical part of the spectrum. 
To this subclass belong most, but not all, BL Lacs discovered 
through their radio emission.
In the second family the synchrotron peak is located at higher 
(UV and soft X--ray) frequencies.
These are most, but not all, radio--selected BL Lacs.
Because of this spectral difference, Padovani \& Giommi (1995a) introduced
the name LBL (low energy BL Lacs) and HBL (high energy BL Lacs)
for the two families.

\begin{figure}
\vskip -2. true cm
\psfig{file=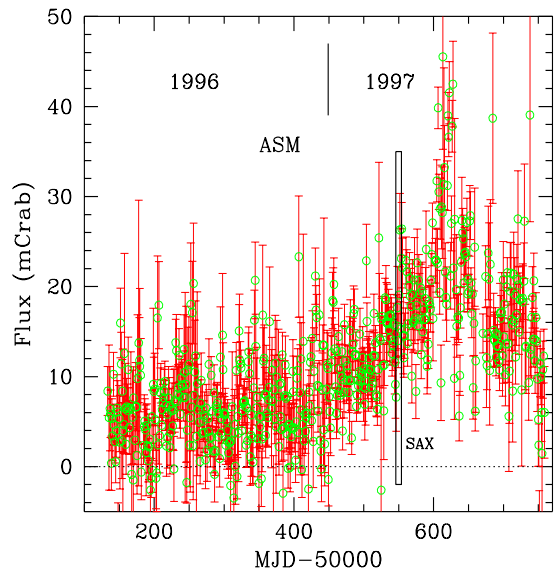,width=8.5truecm,height=9truecm}
\vskip -1 true cm
\caption[h]{All Sky Monitor light curve of Mkn 501. The box
indicates the period of the BeppoSAX observations (7--16 April 1997)}
\vskip -1.5 true cm
\end{figure}

\begin{figure*}
\vskip -1 true cm
\psfig{file=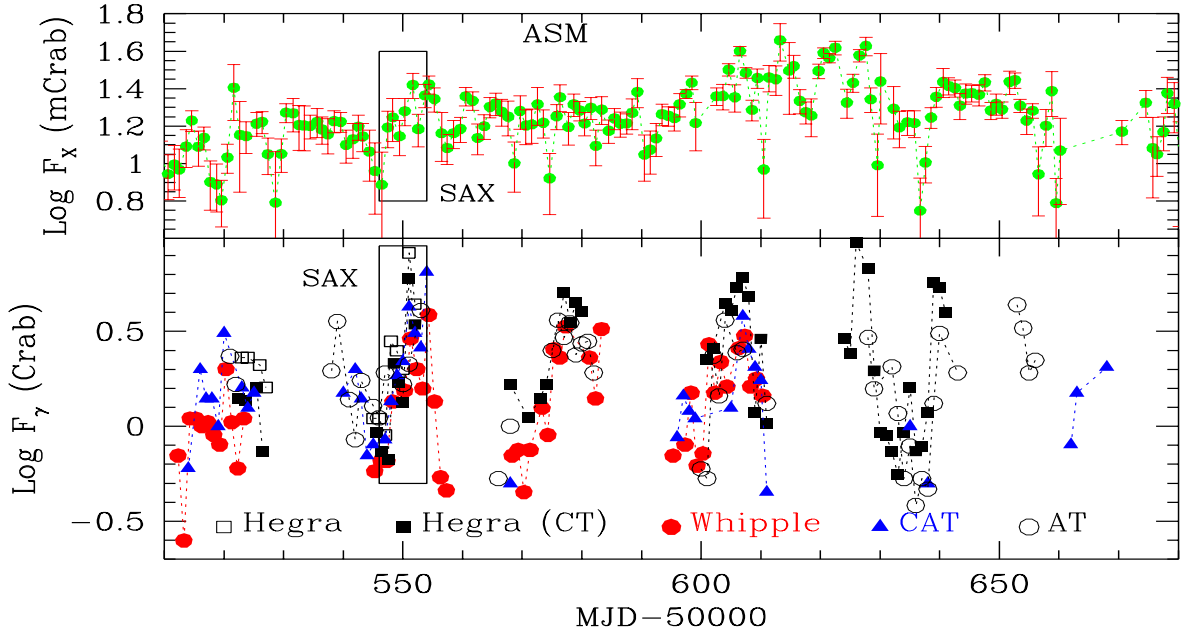,width=17truecm,height=14truecm}
\vskip -5 true cm
\caption[h]{The X--ray (ASM) and TeV light curve of Mkn 501 during 1997.}
\end{figure*}

Mkn 501 is an example of HBL radio--selected BL Lacs, belonging 
to the complete 1 Jy BL Lac and the S4 radio samples, 
but also to the HEAO--1 and the Einstein Slew survey samples
(see e.g. Padovani \& Giommi 1995b).
It is one of the closest BL Lac object ($z=0.034$) and it was the second
BL Lac object, after Mkn 421, to be detected in the TeV energy band 
(Quinn et al. 1996; Bradbury et al. 1997).
Prior to BeppoSAX, it was observed a few times in the X--rays,
showing a spectrum relatively steep (energy index $\alpha_x>1.2$)
in EXOSAT (Sambruna et al. 1994), while in two observations by 
{\it Einstein} the spectrum can be fitted with a power law with index 
flatter than 1, but with large errors (Urry, Mushotzky, \& Holt 1986).
A spectral index $\alpha_x>1$ indicates a synchrotron peak in the far UV region.


According to the All Sky Monitor (ASM) onboard RossiXTE, its X--ray emission
was relatively low during all 1996, but it entered a bright and active
phase from the beginning of 1997, with an average flux roughly twice
the average of the 1996 flux.
This state is still continuing at the time of writing (Dec. 1997).
The ASM light curve is reported in Fig. 1.
During the entire 1997, the source was also extremely active at TeV energies,
as demonstrated by the light curve in Fig. 2, which collects data
of different observatories (Catanese et al. 1997, Aharonian et al. 1997, 
Protheroe et al. 1997) during the period Feb--Sept 1997. 
Note the recurrent rapid flares, during which the source becomes 
a factor 5--10 brighter than the Crab.

BeppoSAX observed Mkn 501 the 7, 11 and 16 of April 1997, 
during one of the TeV flares (Pian et al. 1998). 
Particularly interesting is the last BeppoSAX observation, coincident
with a maximum of the TeV light curve.
The X--ray spectrum revealed by BeppoSAX was exeptional:
it shows the synchrotron peak of its emission at or above 100 keV,
with a brightening of the overall power, with respect to other
previous observations, by a factor $\sim$20.

\section{BEPPOSAX OBSERVATIONS OF MKN 501}
In Fig. 3 we show the LECS+MECS BeppoSAX spectrum data of April 16, 
while in Fig. 4 we show the MECS+PDS spectrum of the same observation.
In Fig. 5 we plot the ratio between the Apr 16 and the Apr 7 spectrum 
(Pian et al. 1998).
This ratio is response matrix and calibration independent.
Note how the spectrum {\it pivots} around $\sim$0.5 keV, and the flattening
of the spectrum in the high (Apr 16) state.
Data analysis confirms this behaviour: in Table 1 we report the relevant
information of the three observations.
The Apr 16 spectrum is well fitted by a broken power law in the LECS+MECS
range, and another broken power law in the MECS+PDS range: confortingly, 
the two separate analysis agree in the intermediate MECS range.
We can therefore conclude that it is likely that the spectrum continuously 
steepens from 0.1 to 100 keV, {\it but remaining always flatter than 
$\alpha_x=1$.}
Therefore, in a $\nu$--$\nu F_{\nu}$ plot, this spectrum peaks
at the highest PDS energies, i.e., above 100 keV.
Note also the large flux in the [13--200] keV band of the 16 Apr 
observations, a factor $\sim$4 larger than 9 days before.
Smaller (30\%) variations in $\sim$3 hours are 
present during each pointing.

\begin{figure}
\vskip -1. true cm
\psfig{file=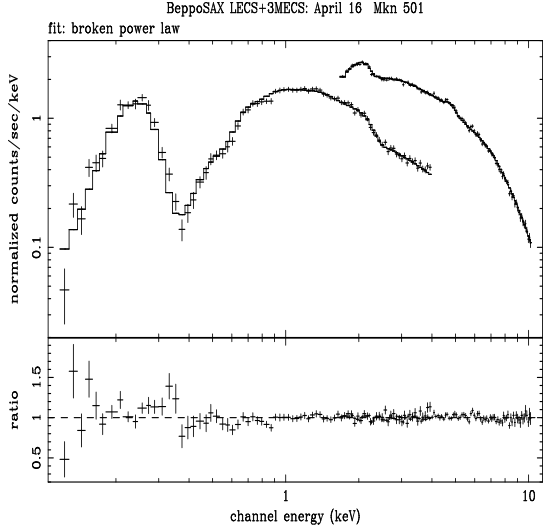,width=9truecm,height=12truecm}
\vskip -5 true cm
\caption[h]{LECS+MECS BeppoSAX spectrum of the April 16 observation
of Mkn 501, fitted with a broken power law.}
\end{figure}
\begin{figure}
\vskip -1 true cm
\psfig{file=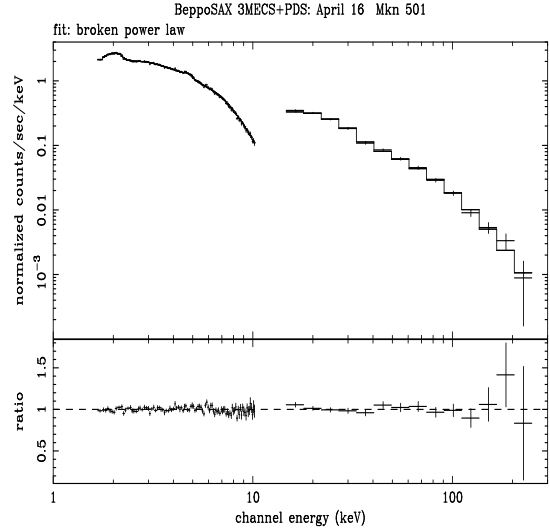,width=9truecm,height=12truecm}
\vskip -5 true cm
\caption[h]{MECS+PDS BeppoSAX spectrum of the April 16 observation
of Mkn 501, fitted with a broken power law. The low energy power law index
is found to perfectly agree with the high energy spectral index of Fig. 3.}
\end{figure}

\begin{figure}
\psfig{file=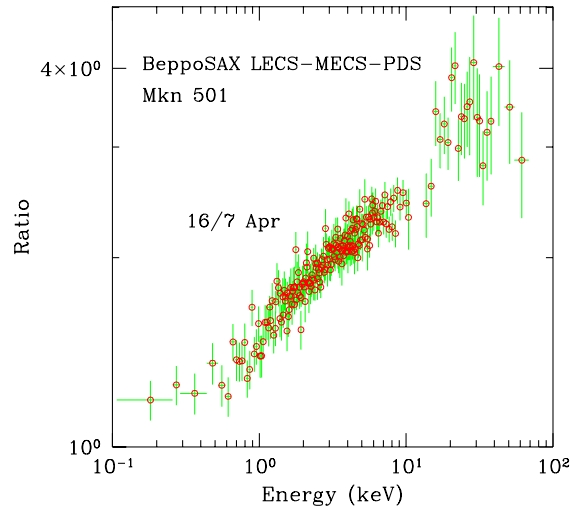,width=8truecm,height=8truecm}
\vskip -1 true cm
\caption[h]{Ratio of the Apr 16 with the Apr 7 BeppoSAX spectra of Mkn 501.
Note the spectral flattening above $\sim$0.5 keV. Below this energy
the spectrum remained almost unchanged.}
\end{figure}

\vskip 0.3 true cm
\begin{table}
\begin{tabular}{lccc} 
\hline
  &    & & \\
  &Apr 7 &Apr 11 &Apr 16 \\
  &    & & \\
\hline
  &    & & \\
$\alpha_1$      &0.63$\pm$ 0.04  &0.64$\pm$0.03 &0.40$\pm$0.03 \\ 
$\alpha_2$      &0.91$\pm$ 0.02  &0.80$\pm$0.02 &0.59$\pm$0.02 \\
$E_b$           &1.76$\pm$ 0.23  &1.85$\pm$0.33 &2.14$^{+0.33}_{-2.14}$ \\
$\alpha_{PDS}$  &0.99$\pm$ 0.13  &0.79$\pm$0.09 &0.84$\pm$0.04 \\
$F_{[2-10]}$    &2.20   &2.45    &5.35 \\
$F_{[13-200]}$  &3.75   &5.15    &18.8 \\
   & & & \\
\hline
  &    & & \\
\end{tabular}
\caption{$\alpha_1$ and $\alpha_2$ are the indices derived by fitting the
LECS+MECS data with a broken power law. 
Unaborbed fluxes in units of $10^{-10}$ erg cm$^{-2}$ s$^{-1}$.}
\end{table}

\section{THE OVERALL SPECTRUM OF MKN 501}
In Fig. 6 we show the overall spectrum of Mkn 501, collecting data which 
are simultaneous or nearly simultaneous with the BeppoSAX observations.
For comparison, we show also a collection of non simultaneous data 
taken from the literature.
Note:

\begin{itemize}

\item The synchrotron peak shifts by almost a factor 1000

\item The X--ray spectrum and flux, below 0.5 keV, do not change
between the BeppoSAX observations

\item According to ISO preliminary data analysis, the far infrared
flux was at the same level of the old IRAS observations (P. Barr,
private communication)

\item Hard X--ray and TeV flux vary together, almost linearly

\item The peak of the Compton component lies at energies below 0.5 TeV

\item With respect with the `normal' state (defined by the ensemble
of the previous non simultaneous data), the source brightened by
a factor $\sim$20, and by a factor $\sim$4 between the 7 and 16 of April.

\end{itemize}

\begin{figure}
\vskip -1 true cm
\psfig{file=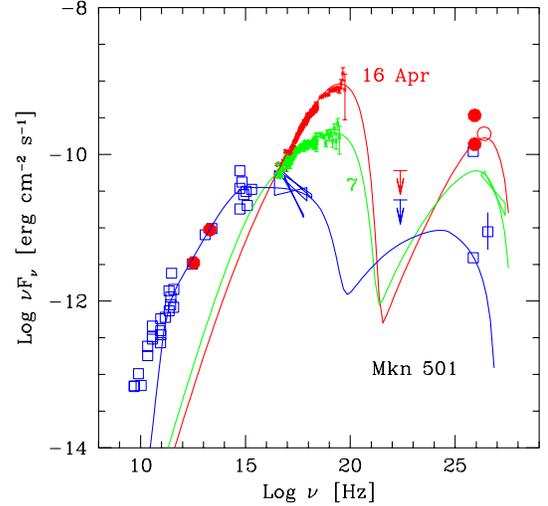,width=8truecm,height=8truecm}
\vskip -1 true cm
\caption[h]{The SED of Mkn 501, adapted from Pian et al. 1998}
\end{figure}

\begin{figure}
\vskip -1 true cm
\psfig{file=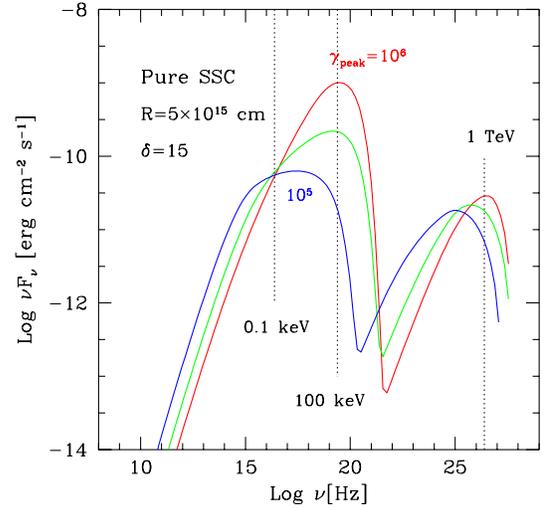,width=8truecm,height=8truecm}
\vskip -1 true cm
\caption[h]{SSC spectra calculated assuming a continuous injection
of relativistic electron through an homogeneous source, for different
values of the injected power.
Note how the self--Compton luminosity increases less than 
than synchrotron luminosity.
}
\end{figure}

\section{MODELLING}

In BL Lac objects the absence of prominent emission lines and 
of any, albeit weak, thermal signature, such as emission from 
an accretion disk or reprocessing due to dust, favours emission 
models based on the synchrotron self--Compton (SSC) emission, where 
ultrarelativistic electrons scatter the synchrotron photons they 
themselves produce (see e.g. Maraschi, Ghisellini \& Celotti 1992). 
On the contrary, in more powerful emitting line blazars, the dominating
process for the formation of the high energy spectrum can be the inverse
Compton scattering of relativistic electrons in the jet off photons
produced outside the jet, for instance by the emission line clouds
or by the accretion disk (EC model, Dermer \& Slickeiser 1992,
Sikora, Begelman \& Rees 1994, 
Ghisellini \& Madau 1996, Ghisellini et al. 1997).

The pure SSC model, however, fails to explain the April observations of Mkn 501.
The main observational constraints are the flattening of the X--ray
spectrum between April 7 and 16, and the simultaneous TeV flux variations.
The radiative cooling time of the electrons producing both the X--ray and
the TeV flux is short, and a continuous resupply of energy must take
place.
This can have the form of a continuous injection of high energy particles,
which find their equilibrium distribution in a time shorter than the
light travel time $R/c$.
In this framework, the observed flattening at X--ray energies is caused
by the flattening of the injected particle distribution.
However, if the particle distribution is flatter at all energies,
the pivoting at $\sim$0.5 keV implies that less particles are
emitting at IR--optical frequencies.
Due to the Klein Nishina decline of the scattering cross section, these
are the relevant target photons for producing the TeV emission.
{\it Even if the high energy electrons increase in number, the target
photons for Compton scattering decrease, resulting in a roughly steady
TeV flux}.
To illustrate this behaviour, Fig. 7 shows the resulting spectra of the
`pure' SSC model where the electron injection varies, becoming flatter
when more powerful, and extending to larger electron energies $\gamma_{max}$.
As can be seen the Compton luminosity is almost constant, even if, in 
restricted energy ranges, the change of $\gamma_{max}$ results
in a possible flux change.
This clearly cannot account for the observed correlated variability.

However the emission, here modelled to come from a single homogeneous zone,
is likely to be produced at a range of radii along a relativistic jet.
Especially so for the radio -- far IR flux.
In addition, even in a simple homogeneous model, it is possible
that one zone of the region undergoes a violent acceleration event
producing a flat electron distribution.
At birth, these electrons are already emdebbed in a `quiescent' 
radiation field, and scatter these {\it ambient} photons together with 
the ones produced by themselves.
The word `ambient' has been chosen to distinguish this model from
the ones where the target photons for Compton scattering are produced
outside the jet.

We have then modelled the spectrum of Mkn 501 along these lines, assuming
that the region producing the X--ray/TeV flare is embedded in a 
{\it steady} radiation field (extending in frequency from the mm to 
the optical), corresponding to the observed flux.

Results of these models are shown in Fig. 6, along with the fit
to the `quiescent' spectrum (i.e. to the collection of non 
simultaneous data prior to the BeppoSAX observations).
Note that in this case we can well reproduced all the main spectral
characteristics of the source.
For all models we assumed a source size $R=5\times 10^{15}$ cm,
a beaming factor $\delta=15$, and a magnetic field $B\sim 0.8$ Gauss.
For the quiescent state, we assumed to continuosly inject relativistic
electrons with a power law distribution $\propto \gamma^{-2}$ between
$\gamma_{min}=3\times 10^3$ and $\gamma_{max}=6\times 10^5$, at a rate
corresponding to an intrinsic injected power of 
$L^\prime=4.6\times 10^{40}$ erg s$^{-1}$.
For the Apr 7 spectrum the injected distribution is $\propto \gamma^{-1.5}$
between $\gamma_{min}=10^4$ and $\gamma_{max}=3\times 10^6$, with 
$L^\prime=1.9\times 10^{41}$ erg s$^{-1}$.
Finally, for the Apr 16 spectrum, the injection is $\propto \gamma^{-1}$ 
between
$\gamma_{min}=4\times 10^5$ and $\gamma_{max}=3\times 10^6$, with
$L^\prime=5.5\times 10^{41}$ erg s$^{-1}$.
According to these parameters, 
the magnetic field energy density is greater
than the particle energy density, but smaller than the overall 
radiation energy density.
On the other hand, if we integrate the radiation energy density
up to the energy $h\nu=1/(\gamma_{max}m_ec^2)$, accounting then for
only those photons available for scattering in the Thomson regime,
we have that the magnetic field is dominant.
Interestingly, the radiative cooling time is equal to the light
crossing time for particles emitting at $\sim 1$ keV, where
there is the pivot in the X-ray spectrum.

\section{CONCLUSIONS}
From the BeppoSAX observations of Mkn 501 we have learned that BL Lacs
can undergo major flaring events, in which their overall, bolometric
power increases by a huge factor, while at the same time increasing
the typical energy of their radiating electrons.
Such events must correspond to a dramatic increase of the power of
the particle acceleration mechanism and of its efficiency in accelerating
electrons up to TeV energies.
The huge shift of the synchrotron peak frequency excludes in fact 
other possibilities, as a change of the beaming factor or of the
magnetic field, since unreasonably large variations would be necessary
in this case $(\nu_{syn} \propto B\delta)$.

We have also learned that it is likely that a pure, simple, one--zone SSC model
cannot account for what we observe, but that there is the need to
invoke another, steadier, source of IR photons as targets for the
inverse Compton process. 
The coincidence of the amount of needed radiation density with what we 
derive from the observed flux is an indication that these photons are 
produced in the vicinity of the hard X--ray and TeV emitting region,
probably by another, steadier, electron population.

Finally we can wonder if Mkn 501 is really exeptional, or if other
sources exist with the same characteristics, namely a synchrotron
peak located in the hard X--rays.
These would garantee the presence of TeV energy electron, and therefore
these sources are the best candidates to be strong TeV emitters.
From what we have observed of Mkn 501, it might be not easy to find
these sources, since: 
i) Mkn 501 was observed with a steep X--ray spectrum prior to BeppoSAX;
ii) even during the April 1997 flare, the low energy X--ray flux
of Mkn 501 was not varying dramatically.

To find these `extreme synchrotron BL Lacs' one should select sources 
with flat ($\alpha_x < 1$) X--ray spectra, checking, by constructing the SED,
if the X--ray spectrum is produced by the synchrotron process 
(in this case the X--rays smoothly connect with the optical-UV data).
For sources without a measured X--ray spectrum, overall spectral indices 
joining the radio, optical and the X--ray band can give some hints, since 
these broad band indices roughly measure the location of the synchrotron peak.

\end{document}